\documentclass[pre,twocolumn,showpacs,superscriptaddress]{revtex4}
\usepackage{amsmath}
\usepackage{bm}
\usepackage{graphicx}
\usepackage{color}

\begin{document}

\title{Glassy dynamics of athermal self-propelled particles: 
Computer simulations and a nonequilibrium microscopic theory}

\author{Grzegorz Szamel}
\affiliation{Department of Chemistry, 
Colorado State University, Fort Collins, CO 80523}

\author{Elijah Flenner}
\affiliation{Department of Chemistry, Colorado 
State University, Fort Collins, CO 80523}

\author{Ludovic Berthier}
\affiliation{Laboratoire Charles Coulomb, UMR 5221 CNRS, 
Universit{\'e} Montpellier, Montpellier, France}

\date{\today}


\begin{abstract}
We combine computer simulations and analytical theory 
to investigate the glassy dynamics in dense assemblies of 
athermal particles evolving under the sole influence of 
self-propulsion. Our simulations reveal that 
when the persistence time of the self-propulsion 
is increased, the local structure becomes 
more pronounced whereas the long-time dynamics first accelerates 
and then slows down. We explain these surprising findings
by constructing a nonequilibrium microscopic theory which 
yields nontrivial predictions for the glassy dynamics. 
These predictions are in qualitative agreement with the simulations and reveal
the importance of steady state correlations of the local velocities
to the nonequilibrium dynamics of dense self-propelled particles. 
\end{abstract} 

\maketitle

\section{Introduction}

The application of statistical mechanical methods to the 
dynamics of individual motile objects started 
shortly \cite{Furth} after Einstein's work 
 on Brownian motion~\cite{Einstein}.
Recently, the \emph{collective} behavior of systems 
consisting of interacting self-propelled particles attracted 
interest \cite{Ramaswamyrev,Marchettirev}. 
An important motivation for studying `active' systems
is to understand spectacular dynamics observed in assemblies of living
systems, such as coherent motion \cite{Bialek,Dombrowski}. 
A more fundamental motivation stems from the nonequilibrium nature 
of active systems that are driven by internal, non-thermal self-propulsion 
forces, which represents a difficult challenge for statistical physics. 
The behavior of such systems may defy our equilibrium-based
physical intuition, as demonstrated by large-scale collective
motion in persistent hard disk systems~\cite{Deseigne}, and the emergence of
dynamic clustering~\cite{Theurkauff,Buttinoni}, phase 
separation~\cite{Buttinoni,Redner,Cates}, and nonequilibrium 
equation of state~\cite{Ginot,Takatori,Solon} in repulsive self-propelled particles.

Active particles may also
exhibit behavior similar to that found in equilibrium systems, 
such as crystallization \cite{Bialke1}. This behavior suggests that, like 
thermal systems \cite{BerthierBiroli},
dense active systems could possibly be supercooled and 
exhibit glassy dynamics; as recently argued theoretically 
on the basis of a simple 
driven glassy model~\cite{BerthierKurchan}. This
study was followed by numerical investigations of active particles
with hard-core \cite{Ni,Berthier} or continuous \cite{Gompper,Dasgupta} 
interactions. Active glassy dynamics was observed, but when compared with 
thermal systems, the onset of glassy behavior was always pushed towards 
higher densities and lower temperatures.
When self-propulsion is progressively added to an otherwise 
thermal system, it was noted that the local structure
becomes less pronounced \cite{Ni,Dasgupta}. Such change in local structure
suggests both a simple physical explanation for the shift of the onset
of glassy behavior, and that a straightforward extension of mode-coupling 
theory \cite{Goetzebook} can describe this 
shift \cite{BerthierKurchan,Farage}. 

Here we present a simulational and theoretical study of the structure 
and glassy dynamics of a more complex system in which 
{\it self-propulsion is the only source of motion}. 
We study self-propelled particles interacting with a continuous potential 
(differently from \cite{Ni,Berthier}) and without thermal Brownian motion 
(differently from \cite{Ni,Bialke1,Gompper}). Thus, our model
is `athermal'~\cite{Fily,Berthier}, and  
the degree of nonequilibrium is quantified by the persistence time 
$\tau_p$ of the self-propulsion. As $\tau_p$ increases 
at a constant effective temperature, we find 
that the structure of the system
becomes more pronounced, whereas the dynamics initially speeds up and 
then slows down, showing that in our model the shift of the glassy dynamics 
with self-propulsion is not simply the direct consequence 
of the changing microstructure \cite{Farage,Dasgupta}.
 
To elucidate these findings we develop a microscopic theory 
that accounts for the nonequilibrium nature of 
athermal self-propelled systems. We are aware of no other approach
where many-body interactions are taken into account at the microscopic level. 
The theory is constructed from the steady state 
structure factor and the steady state correlations of the velocities. 
The latter correlations are non-trivial only for self-propelled systems 
with a finite persistence time, and are central to explain 
the opposite, and seemingly contradictory, evolution of the structure and 
dynamics in dense active materials.

\section{Model active system}

We study a system of interacting 
self-propelled particles in a viscous medium.
We model self-propulsion as an internal driving
force evolving according to the 
Ornstein-Uhlenbeck \cite{VanKampen} process: 
\begin{eqnarray}\label{eompos}
\dot{\mathbf{r}}_i &=& \xi_0^{-1}\left[\mathbf{F}_i  + \mathbf{f}_i
\right], \\ \label{eomsp}
\dot{\mathbf{f}}_i &=& -\tau_p^{-1} \mathbf{f}_i + \boldsymbol{\eta}_i.
\end{eqnarray}
In Eq.~(\ref{eompos}), $\mathbf{r}_i$ is the position of particle $i$,
$\xi_0$ the friction coefficient of an isolated particle,
$\mathbf{F}_i$ is the force acting on particle $i$ originating 
from interactions, 
$\mathbf{F}_i = -\sum_{j\neq i} \boldsymbol{\nabla}_i V(r_{ij})$,
and $\mathbf{f}_i$ is the self-propulsion acting on particle $i$.  
In Eq. (\ref{eomsp}), 
$\tau_p$ is the persistence time of the self-propulsion and
$\boldsymbol{\eta}_i$ is an internal Gaussian noise with zero mean and 
variance $\left<\boldsymbol{\eta}_i(t) \boldsymbol{\eta}_j(t')
\right>_{\text{noise}} = 
2 D_f\boldsymbol{I}\delta_{ij}\delta(t-t')$, where $\left< ... 
\right>_{\text{noise}}$ 
denotes averaging over the noise distribution, $D_f$ is the noise strength
and $\boldsymbol{I}$ is the unit tensor.
Without interactions, Eqs.~(\ref{eompos}-\ref{eomsp}) 
produce a persistent random walk with a self-diffusion coefficient 
$D_{0} = D_f\tau_p^2/\xi_0^2$, which defines the  
\textit{single-particle} effective temperature: 
$T_{\text{eff}}= D_{0} \xi_0  = D_f \tau_p^2/\xi_0$  \cite{toy}.
It is convenient to choose as three independent control 
parameters the number density $\rho$ (which is kept fixed), the effective 
temperature $T_{\text{eff}}$, and the persistence time $\tau_p$. The 
persistence time quantifies the degree of nonequilibrium; when $\tau_p 
\to 0$ our system becomes equivalent to a Brownian system 
at temperature $T=T_{\text{eff}}$. Including additional thermal noise would 
add a fourth control parameter to the model. Finally, we should mention that recently 
an approximate mapping has been derived  \cite{FarageKB} 
between our model and the standard 
active Brownian particles model \cite{tenHagen}.

\section{Computer simulation study}

\begin{figure}
\includegraphics[width=1.5in]{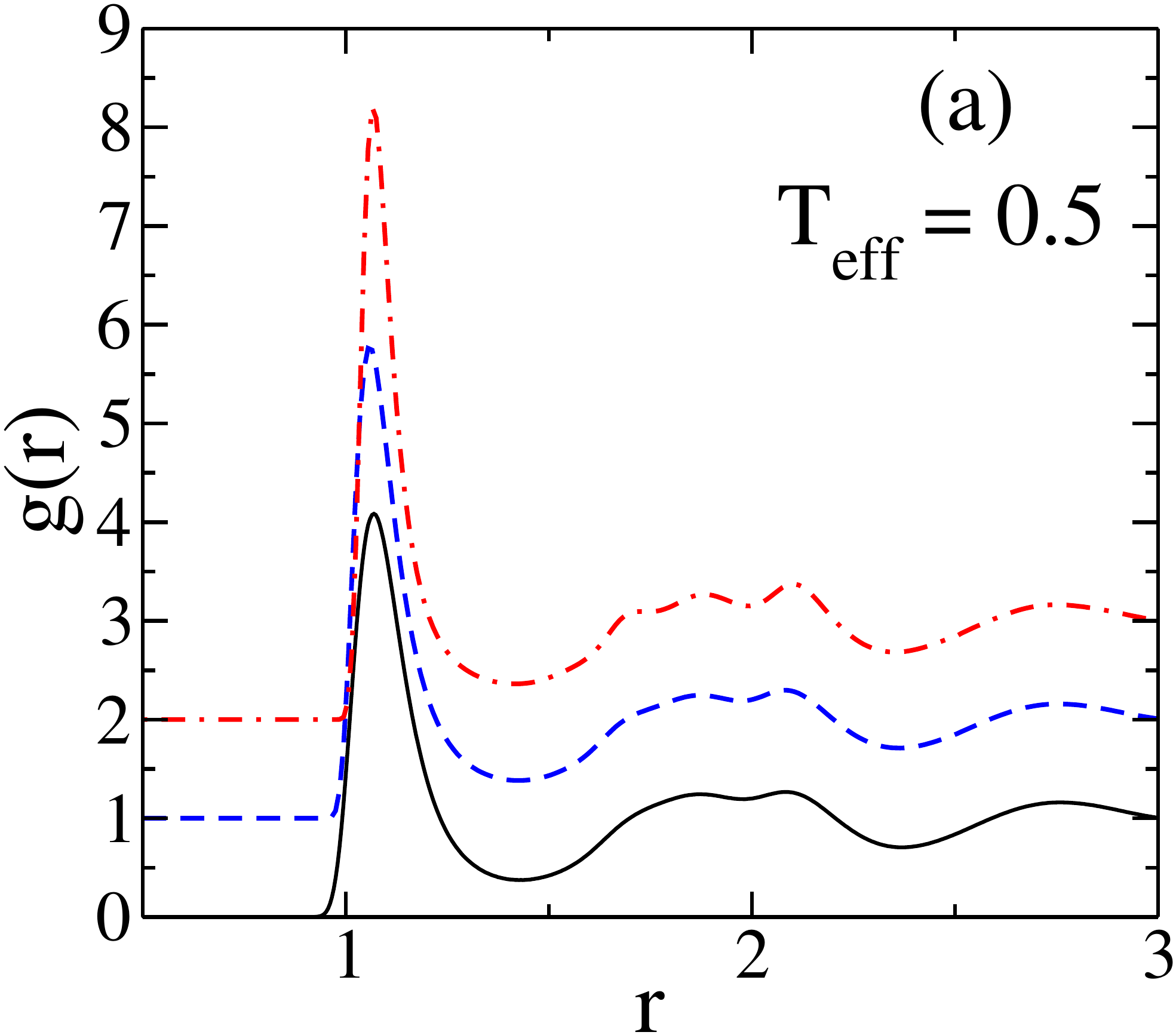}\hskip 1em
\includegraphics[width=1.5in]{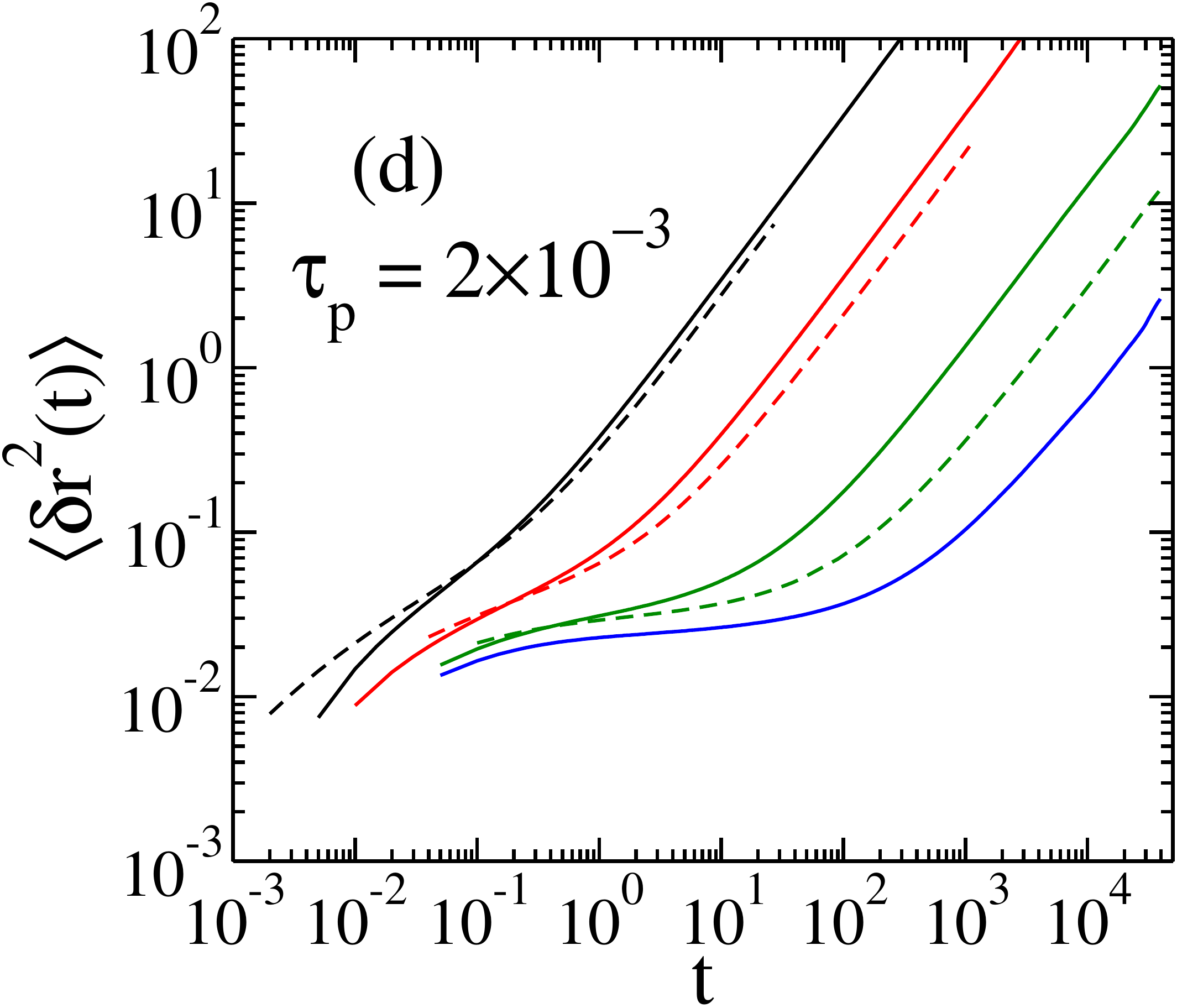}\\[1ex]
\includegraphics[width=1.5in]{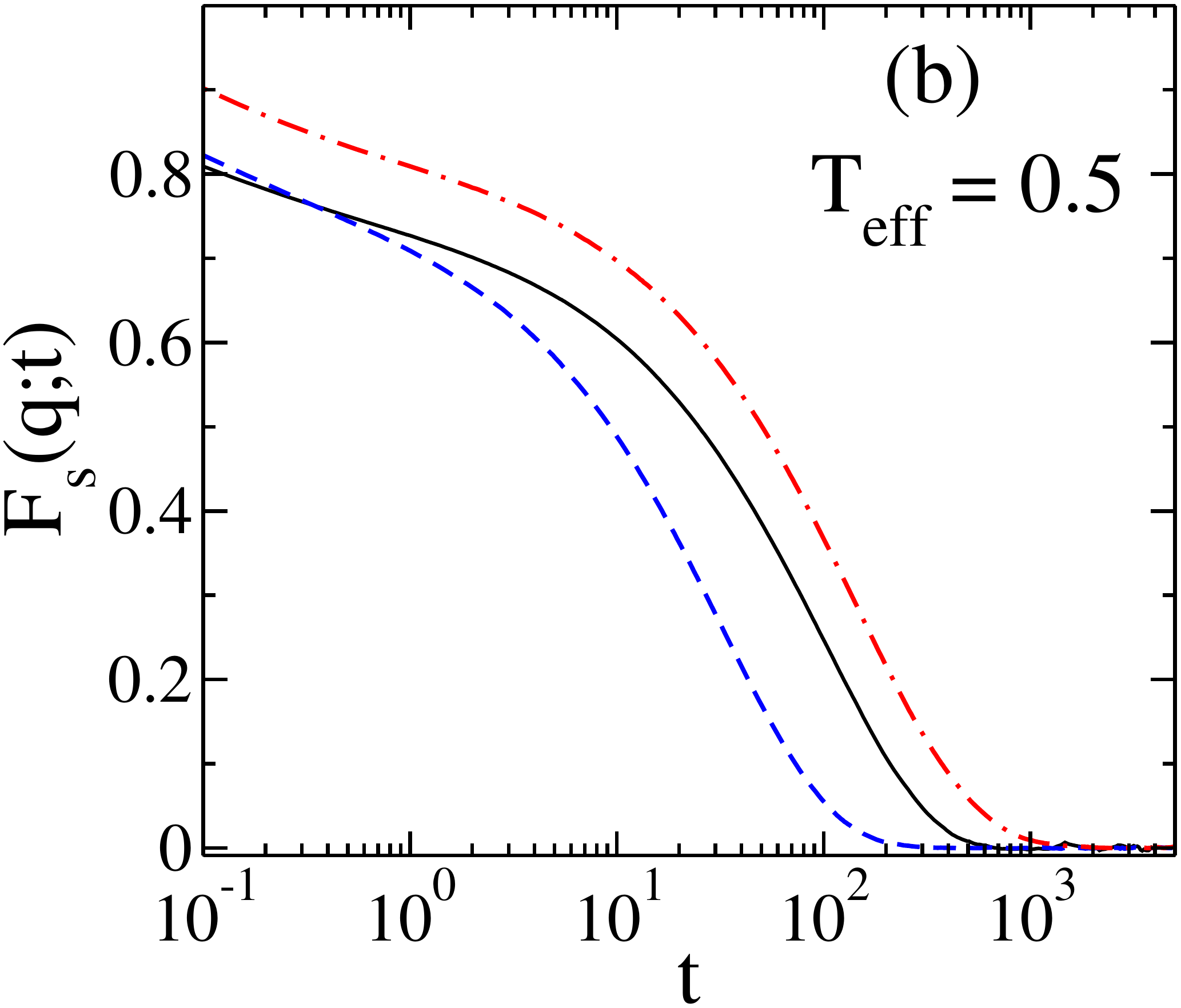}\hskip 1em
\includegraphics[width=1.5in]{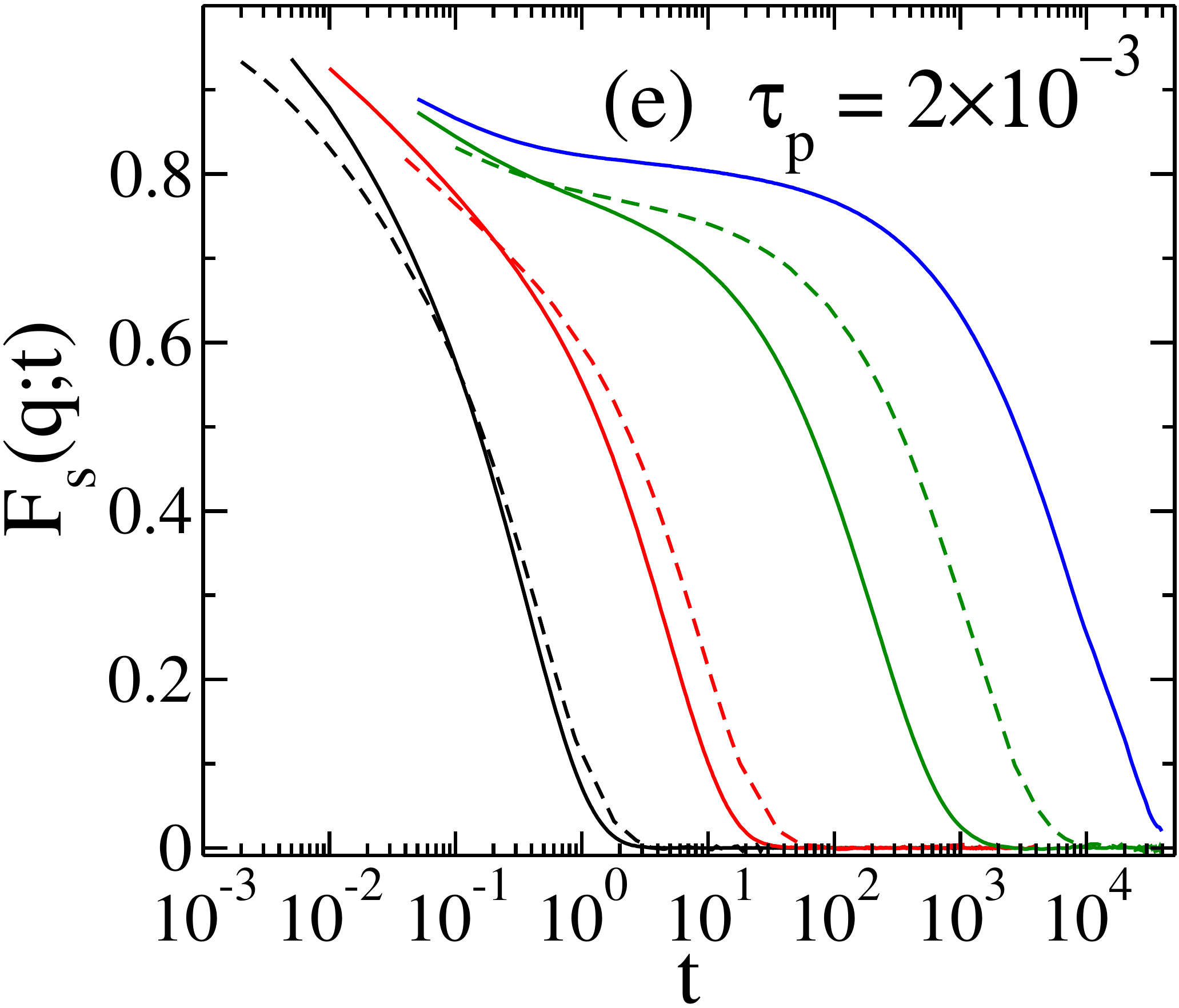}\\[1ex]
\includegraphics[width=1.5in]{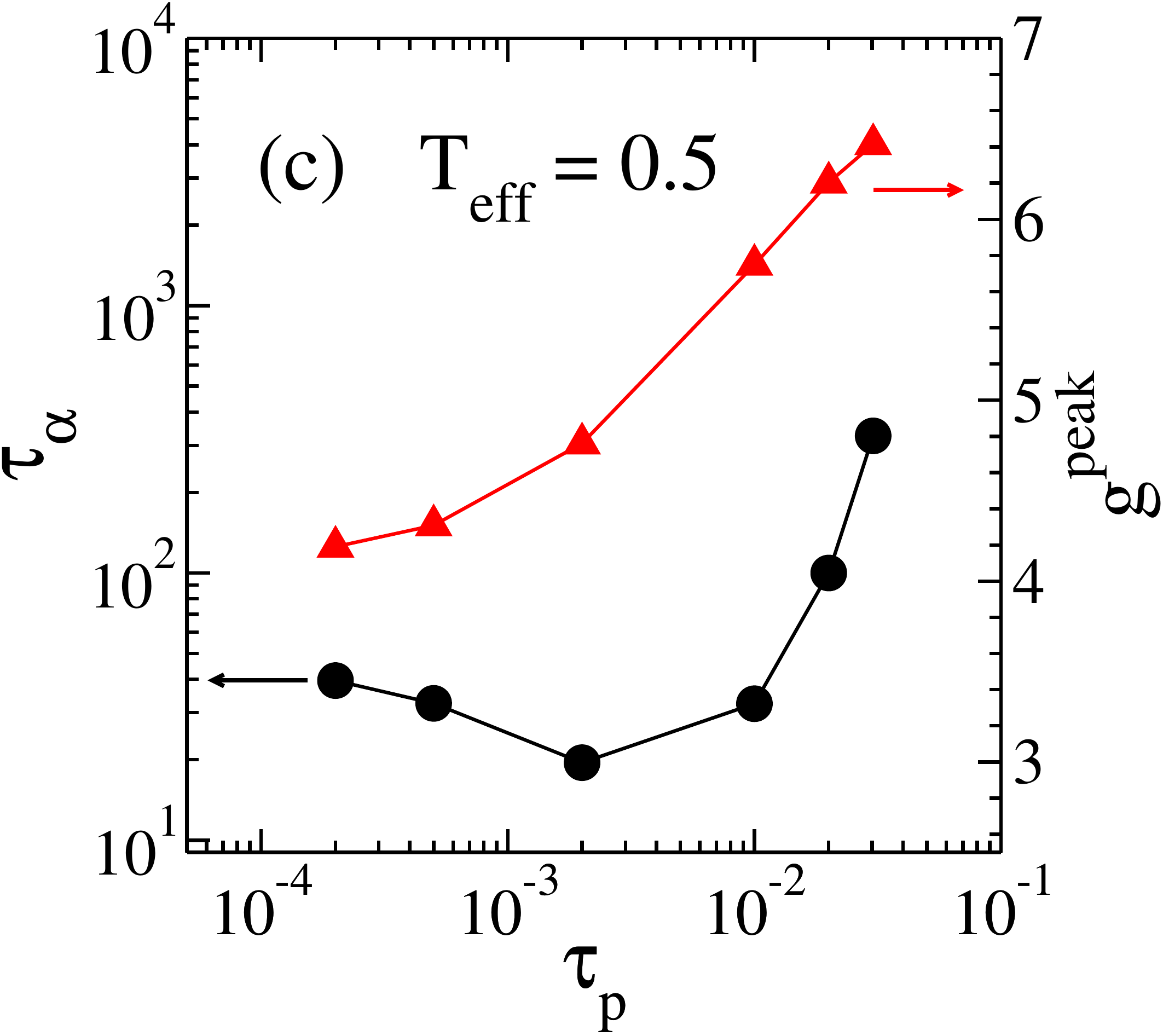}\hskip 1em
\includegraphics[width=1.5in]{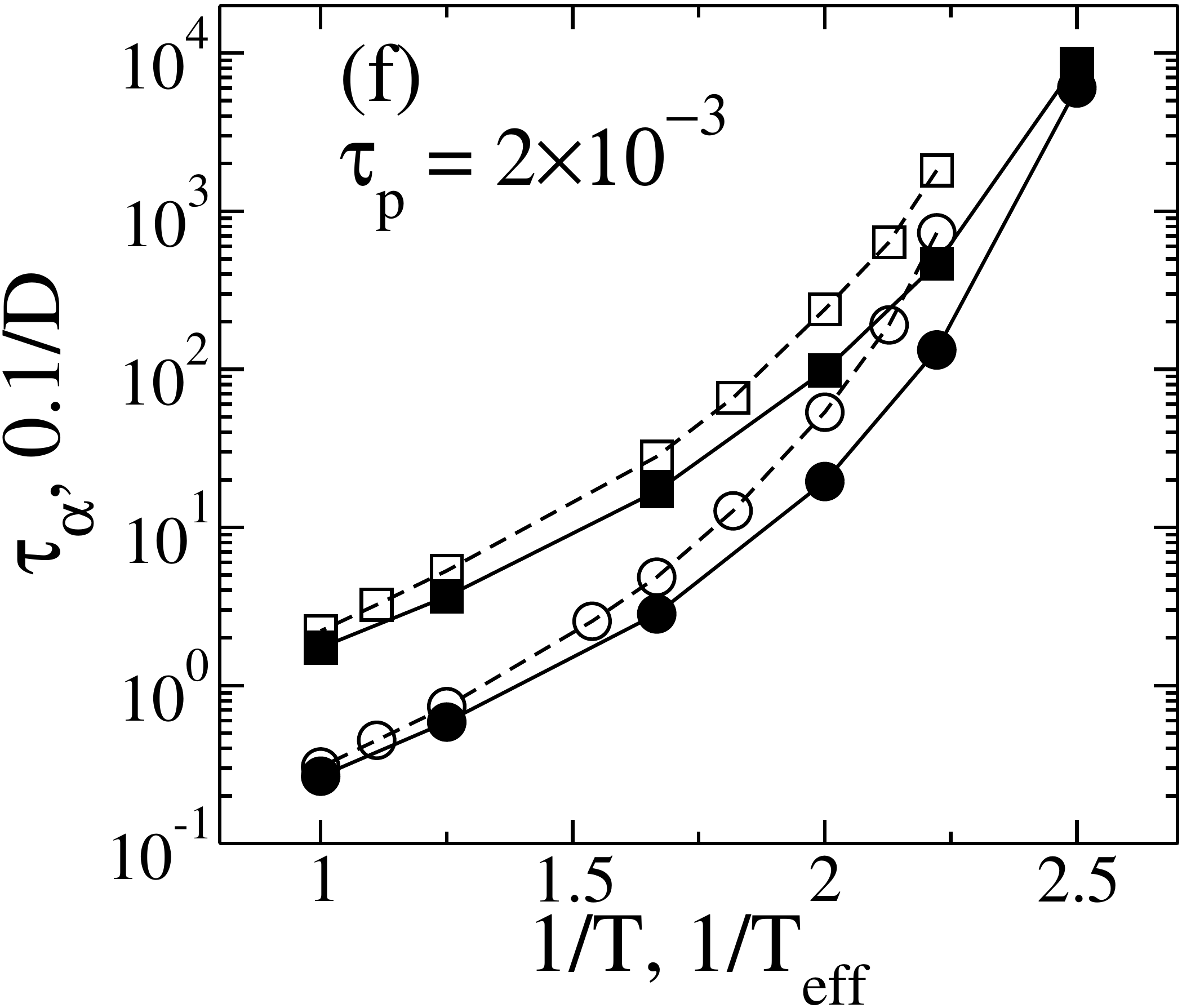}
\caption{\label{fig1}
Evolution of the structure and dynamics with
the persistence time $\tau_p$ at $T_{\text{eff}}=0.5$ (a-c) and
comparison of dynamics in active systems with $\tau_p=2\times 10^{-3}$ 
and Brownian systems at $T=T_{\rm eff}=0.5$ (d-f). 
In (a,b) solid lines correspond to $\tau_p=0$ 
(equivalent to a thermal system at $T=0.5$),
and the dashed, and dot-dashed lines correspond to 
$\tau_p=2\times 10^{-3}$, and $2\times 10^{-2}$, respectively. 
In (a) we show the pair distribution function $g(r)$; the curves are shifted
vertically for clarity. In (b) we show the self-intermediate 
scattering function
$F_s(q;t)=\langle e^{i \mathbf{q} \cdot (\mathbf{r}_j(t)-\mathbf{r}_j(0))}\rangle$ 
for $q=7.25$. In (c) we show
the relaxation time, $\tau_\alpha$ (circles) and the 
peak value of $g(r)$ (triangles). The relaxation time
is defined as $F_s(q;\tau_\alpha)=e^{-1}$. 
In (d,e) solid lines correspond
to active systems at $T_{\text{eff}}=1$, 0.6, 0.45 and 0.4, and dashed lines 
correspond to equilibrium systems at $T=1$, 0.6 and 0.45 
(from left to right). 
In (d) we show mean-square displacement $\left<\delta r^2(t)\right> = 
\left<(\mathbf{r}_j(t)-\mathbf{r}_j(0))^2\right>$ and in (e) we show 
$F_s(q;t)$. In (f) circles represent $\tau_\alpha$, squares
represent the inverse self-diffusion coefficient, $0.1/D$
of active (closed symbols) and thermal (open symbols) systems.}
\end{figure}

To compare the glassy behavior of the 
self-propelled system with that of a well-studied thermal system,
we simulated the Kob-Andersen (KA) 
binary mixture \cite{Kob1994}. All quantities presented 
pertain to the large particles, which comprise 80\% of the mixture. The results
are presented in reduced units \cite{units} at the well-studied 
density $\rho=1.2$.  
In panels (a-c) of Fig. \ref{fig1} we show the dependence of the structure
and dynamics when moving 
away from equilibrium by increasing  
the persistence time of the self-propulsion at constant 
$T_{\text{eff}}=0.5$~\cite{commentT}. At $T=0.5$  
the thermal KA system exhibits glassy dynamics. In panels (a,c),
we see that the pair correlation $g(r)$ of the 
active fluid becomes more structured at all length 
scales with increasing $\tau_p$.
In equilibrium, such behavior is usually accompanied by 
slower dynamics. Surprisingly, panels (b,c) show that 
the nonequilibrium dynamics exhibits a non-monotonic 
dependence on $\tau_p$, which allows one to
define an `optimal' value for $\tau_p$. Describing the contrasting 
dependencies of structure and dynamics is a theoretical challenge
since most microscopic glass theories predict the dynamics on 
the basis of the pair structure~\cite{BerthierBiroli,Gilles}.
A reentrant behavior of the dynamics in driven glassy dynamics
is typically not observed \cite{BerthierKurchan,Ni,Berthier,Gompper,Dasgupta}.
In panels (e,f) we show that, if the persistence time is chosen 
at its optimal value, the dependence of the dynamics of the 
self-propelled system on $T_{\rm eff}$ is weaker than that of the 
thermal Brownian system on $T$. 
In particular, we can study the self-propelled system
at $T_{\text{eff}}=0.4$, whereas it is challenging to simulate 
the thermal system below $T \approx 0.43$. 
The opposite is true for longer persistence times, where 
the dependence of the dynamics of the self-propelled system 
on $T_{\rm eff}$ becomes significantly more pronounced than that of 
the thermal Brownian system.

\section{Theory}

We now outline a microscopic theory for the time 
dependence of the collective intermediate scattering 
function, $F(q;t)$, of our model active system~\cite{commentFs},
\begin{equation}\label{Fkt}
F(q;t)=N^{-1}\left<n(\mathbf{q})e^{\Omega t} n(-\mathbf{q})
\right>.
\end{equation}
In Eq.~(\ref{Fkt}), $N$ is the number of particles, 
$n(\mathbf{q})  = \sum_j e^{-i\mathbf{q}\cdot\mathbf{r}_j}$ 
is the Fourier transform of the microscopic density,
and $\Omega$ is the $N$-particle evolution operator that can be derived
from the equations of motion (\ref{eompos}-\ref{eomsp}):
\begin{equation}\label{Omega1}
\Omega = -\xi_0^{-1} \sum_{i} \boldsymbol{\nabla}_i\cdot 
\left(\mathbf{f}_i + \mathbf{F}_i \right) 
+ \sum_{i} \frac{\partial}{\partial \mathbf{f}_i}\cdot
\left( \tau_p^{-1} \mathbf{f}_i  + D_f \frac{\partial}{\partial 
\mathbf{f}_i}\right).
\end{equation}
Finally $\left<\dots\right>$ in (\ref{Fkt}) denotes
an average over a steady state distribution 
of positions and self-propulsions; 
the steady state distribution stands to the right of the 
quantity being averaged, and all operators act on it too.
In our approach, we first integrate out the self-propulsions
and then we use the projection operator method, and
a mode-coupling-like approximation to derive 
an approximate equation of motion for $F(q;t)$. 

To begin, we briefly discuss the case of 
non-interacting particles. In this case one could start from the 
Laplace transform of Eq.~(\ref{Fkt}) with the evolution operator similar to 
(\ref{Omega1}) but without interactions. In the simplest 
approximation, after integration over self-propulsions one gets
\begin{eqnarray}\label{nonint}
F(q;z)=N^{-1}\left<n(\mathbf{q}) 
\left(z-\Omega^{\text{eff}}_{\text{free}}(z)\right)^{-1} 
n(-\mathbf{q})\right>_{\bf r}, 
\end{eqnarray}
where $\Omega^{\text{eff}}_{\text{free}}(z) = 
\xi_0^{-2} \sum_i \boldsymbol{\nabla}_i \left(z+\tau_p^{-1}\right)^{-1}
D_f \tau_p \cdot \boldsymbol{\nabla}_i$ and $\left< ... \right>_{\bf r}$ denotes 
the steady state average over particles positions.
According to 
$\Omega^{\text{eff}}_{\text{free}}(z)$, particle motion is ballistic at
short times and diffusive at long times, with the 
long-time self-diffusion coefficient $D_{0}$ discussed 
above. 

For \emph{interacting} particles, the integration over 
self-propulsions is more complicated due to 
non-trivial correlations between positions and self-propulsions 
(already present for a single self-propelled particle in 
an external potential \cite{toy}).
As we show in Appendix A, the final result is a formula analogous 
to Eq.~(\ref{nonint}) but with the following evolution operator,
\begin{eqnarray}\label{Omega2}
\lefteqn{\Omega^{\text{eff}}(z) = \xi_0^{-2} \sum_{i,j} 
\boldsymbol{\nabla}_i \cdot \left(z+\tau_p^{-1}\right)^{-1} }
\nonumber \\ &&
\left(\left<\mathbf{f}_i \mathbf{f}_j\right>_{\text{lss}} 
- \left<\mathbf{f}_i\right>_{\text{lss}} \left<\mathbf{f}_j
\right>_{\text{lss}}\right)
\cdot
\left[-\mathbf{F}_j^{\text{\text{eff}}}+\boldsymbol{\nabla}_j\right].
\end{eqnarray}
Here, $\left< ... \right>_{\text{lss}}$ is an average over the conditional steady
state distribution of self-propulsions, 
$P_N^{\text{ss}}(\mathbf{r}_1,\mathbf{f}_1, ..., \mathbf{r}_N,\mathbf{f}_N)/
P_N^{\text{ss}}(\mathbf{r}_1, ..., \mathbf{r}_N)$, where the superscript 
`$\text{ss}$' stands for `steady state'. Furthermore, 
$\mathbf{F}_j^{\text{\text{eff}}} = 
\boldsymbol{\nabla}_j \ln P_N^{\text{ss}}(\mathbf{r}_1, ..., \mathbf{r}_N)
$ is the effective force
acting on particle $j$ in the steady state.
The most important physical assumption used in the derivation of Eq.~(\ref{Omega2}) is 
the absence of systematic currents in the steady state, see Appendix A.
We expect the persistence time to be renormalized by the 
interactions; the presence of the bare 
persistence time in Eq. (\ref{Omega2}) represents an approximation.  

We now use the projection operator method and arrive at the following 
memory function equation,
\begin{eqnarray}\label{mfe}
\lefteqn{
\partial^2_t F(q;t) +\tau_p^{-1} \partial_t  F(q;t) =  }
\\ \nonumber &&  
-\frac{\omega_{\parallel}(q) q^2}{S(q)} F(q;t)
- \int_0^t dt' M^{\text{irr}}(q;t-t') \partial_{t'} F(q;t').
\end{eqnarray}
Here, $S(q)=\left< n(\mathbf{q}) n(-\mathbf{q})\right>$ 
is the \textit{steady state} structure factor, 
$\omega_{\parallel}(q) = \left(N\xi_0^2\right)^{-1}
\left< \left|\hat{\mathbf{q}} \cdot
\sum_i \left(\mathbf{f}_i+\mathbf{F}_i\right)
e^{-i\mathbf{q}\cdot\mathbf{r}_i}\right|^2 \right>$ quantifies correlations 
of the velocities of individual particles~\cite{commentv},
and $M^{\text{irr}}(q;t)$ is the irreducible memory function.
The presence of the second time derivative in Eq.~(\ref{mfe})
reflects the ballistic nature of the short-time motion.
A comparison of Eq.~(\ref{mfe}) with the analogous equation for 
an underdamped thermal system suggests to interpret 
$\tau_p \omega_{\parallel}(q)/S(q)$ as a short-time 
collective diffusion coefficient. Since this coefficient 
involves $\omega_{\parallel}(q)$, even in the absence of the memory function 
we need \textit{two} static correlation functions to predict 
the dynamics, $S(q)$ and $\omega_{\parallel}(q)$. Whereas the 
emergence of velocity correlations can be generically 
expected far from equilibrium, the specific role they play for 
self-propelled systems is non-trivial and was not identified before.

The main approximation of our theory is 
a factorization approximation for the memory function
to close the dynamical equations. This is analogous to the
mode-coupling approximation~\cite{Goetzebook}. As we show in Appendix B, 
using the factorization approximation and an approximation for the steady state 
vertex function, we arrive at the following expression for the
memory function,
\begin{eqnarray}\label{memfction}
M^{\mathrm{irr}}(q;t) &=&
\frac{\rho 
\omega_{\parallel}(q)}{2} \int \frac{d\mathbf{q}_1 d\mathbf{q}_2}{(2\pi)^3}
\delta(\mathbf{q}-\mathbf{q}_1-\mathbf{q}_2) \\ \nonumber && \times
\left(\hat{\mathbf{q}}\cdot\left[ \mathbf{q}_1 \mathcal{C}(q_1) + 
\mathbf{q}_2 \mathcal{C}(q_2)
\right]\right)^2 F(q_1;t)F(q_2;t).
\end{eqnarray}
Equation~(\ref{memfction}) has a structure similar to the 
memory function of the mode-coupling theory, but it involves a new function 
$\mathcal{C}(q)$ (which replaces the direct correlation function $c(q)$
in the mode-coupling $M^{\text{irr}}(q;t)$),
\begin{eqnarray}\label{vertex}
\displaystyle \rho \mathcal{C}(q) = 1-\frac{\omega_{\parallel}(q)}
{\omega_{\parallel}(\infty )S(q)},
\end{eqnarray}
where
$\omega_{\parallel}(\infty ) =  \left(3N\xi_0^2\right)^{-1} 
\left<\sum_i \left(\mathbf{f}_i
+\mathbf{F}_i\right)^2\right>$.
Equations (\ref{mfe}-\ref{vertex}) are closed and
can be solved if static steady state functions 
$S(q)$ and $\omega_{\parallel}(q)$ are available. To test 
the theory quantitatively, we used the 
static information obtained directly from simulations. Using the
KA system would require formulating and solving 
our theory for a binary mixture. Instead, we 
performed additional simulations of a one-component Lennard-Jones (LJ) 
system of self-propelled particles~\cite{commentLJ}.
We measured $S(q)$ and $\omega_{\parallel}(q)$ and used them to solve 
Eqs.~(\ref{mfe}-\ref{vertex}) numerically to compare the dynamics 
predicted by the theory with the numerical results. 
We focus on the observed non-monotonic evolution of the dynamics 
with the persistence time because it represents a demanding test
of the theory.

\begin{figure}
\includegraphics[width=1.5in]{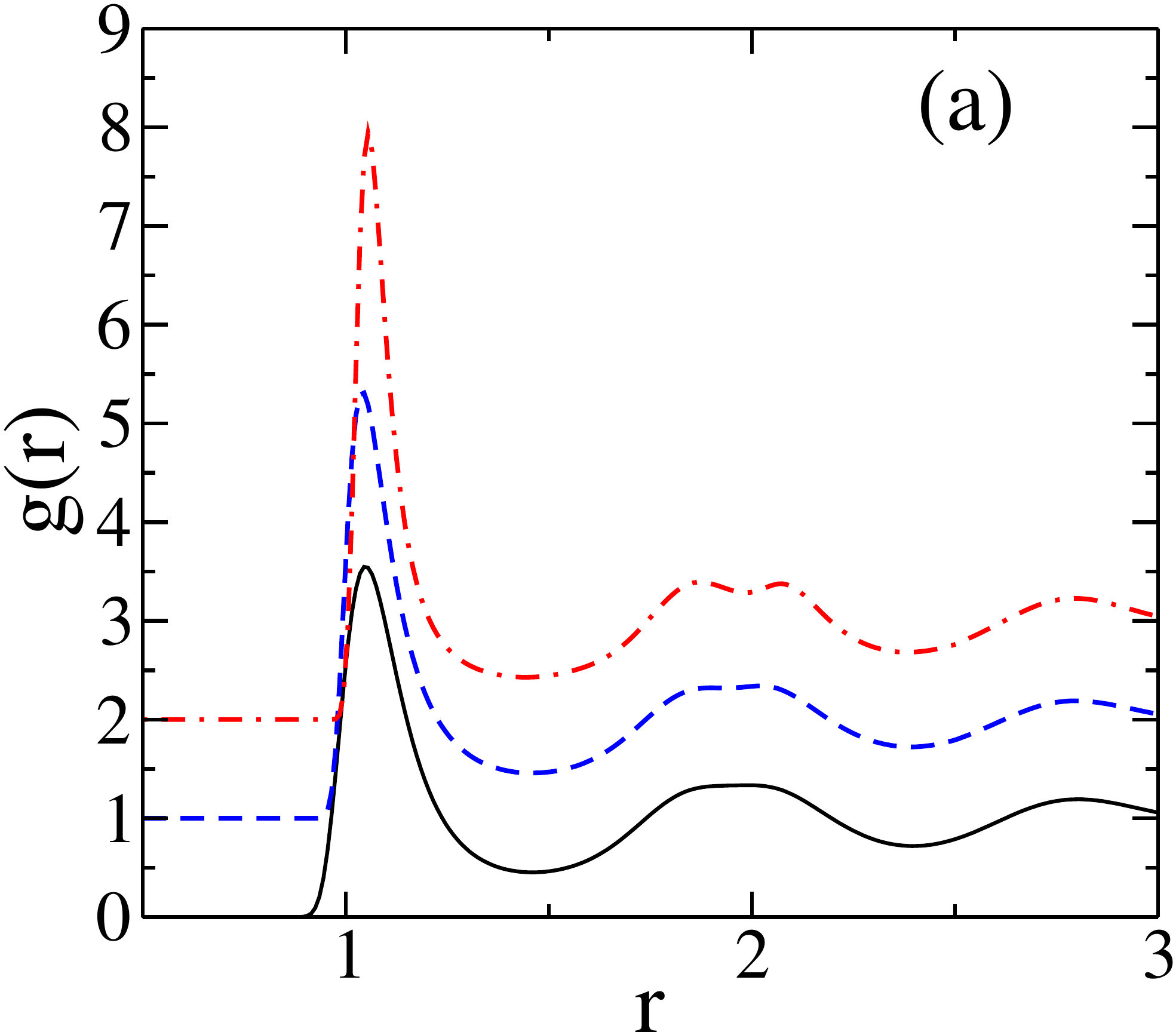}\hskip 1em
\includegraphics[width=1.5in]{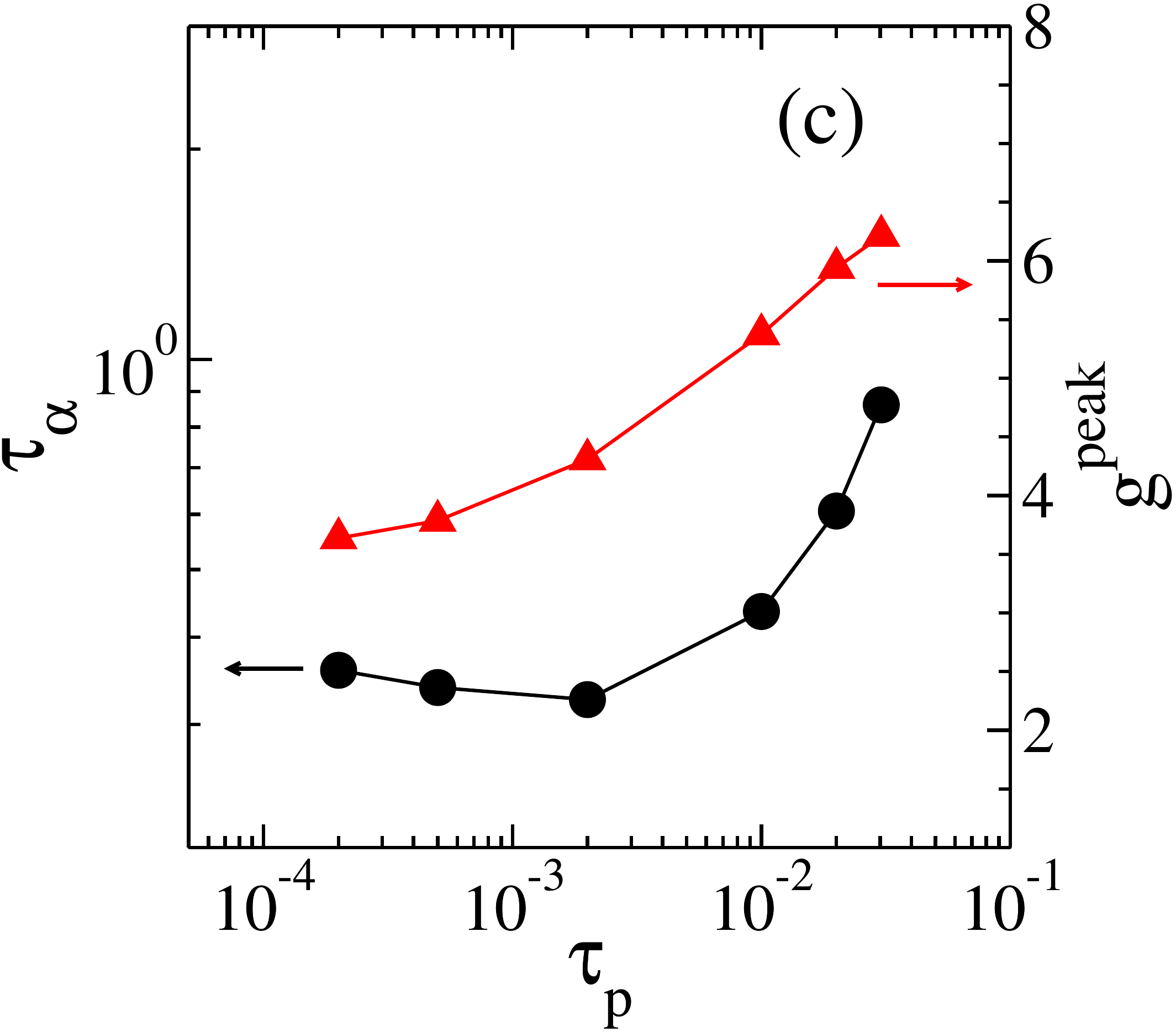}\\[1ex]
\includegraphics[width=1.5in]{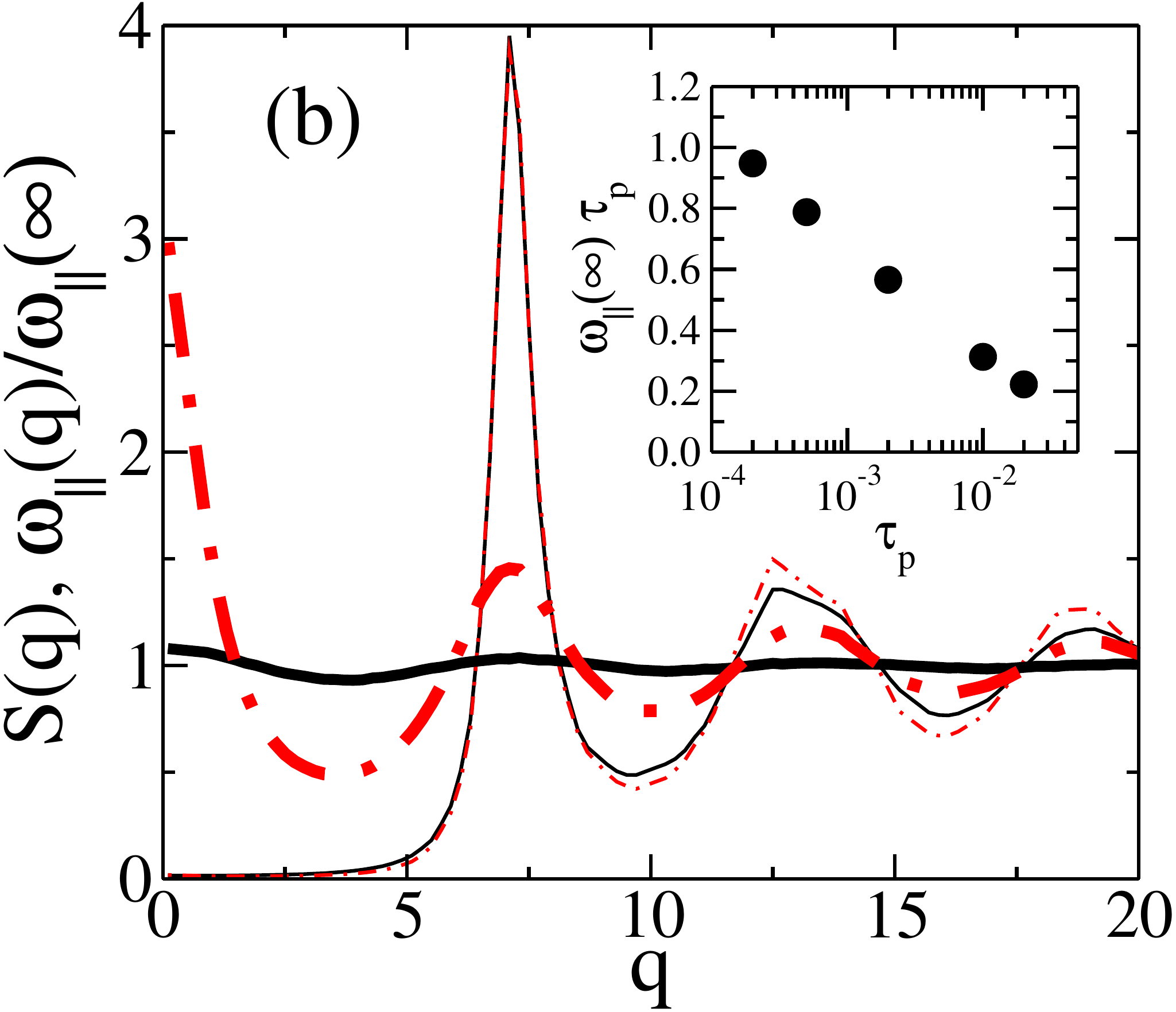}\hskip 1em
\includegraphics[width=1.5in]{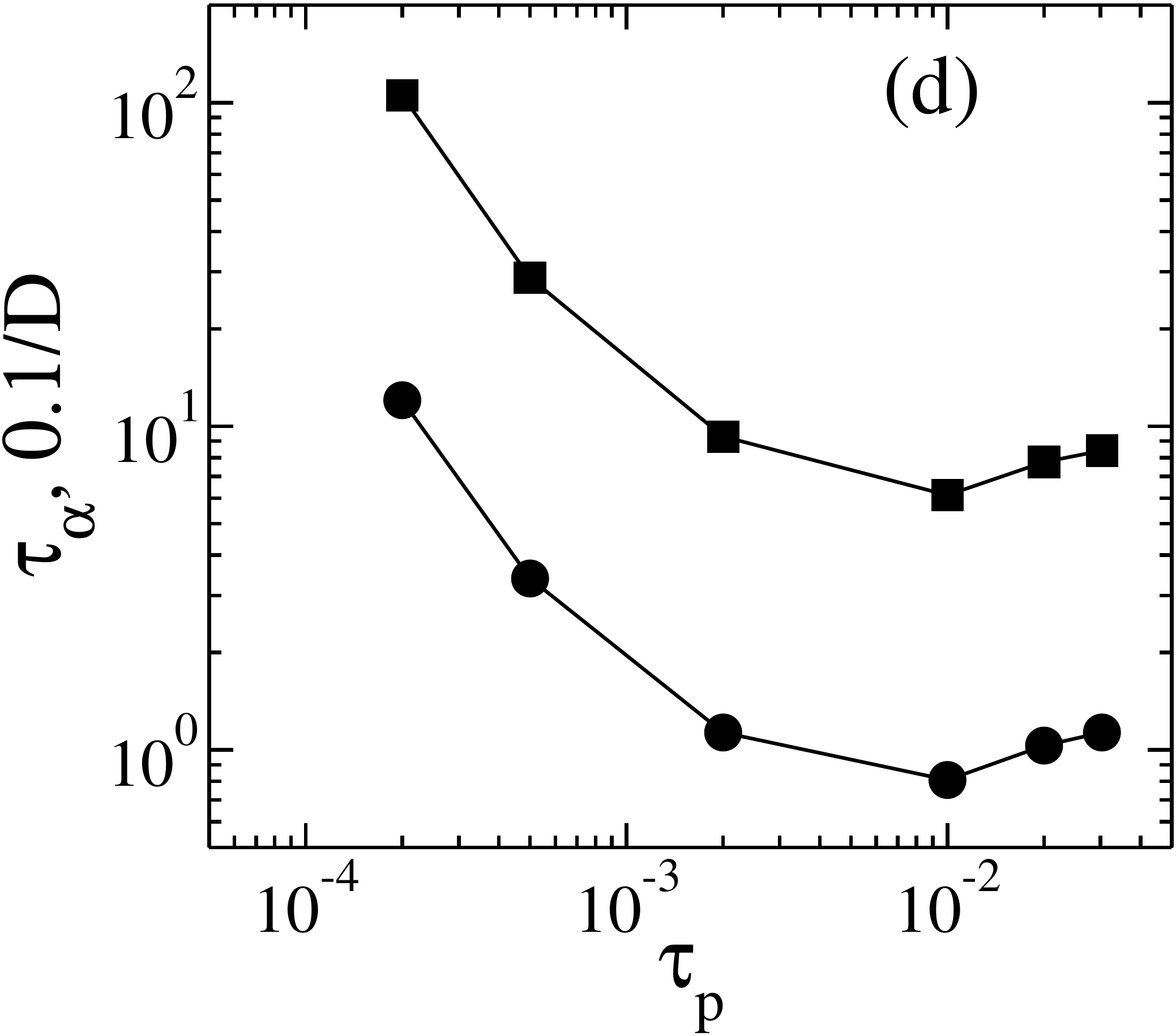}
\caption{\label{fig2}
Structure (a,b) and dynamics (c) obtained from simulations, and predicted by
the theory (d) for various
$\tau_p$ at $T_{\text{eff}}=0.9$ in the one-component LJ system. 
(a) Pair distribution function $g(r)$; the curves are shifted
vertically for clarity; $\tau_p=0, 2 \times 10^{-3}$, and 
$2\times 10^{-2}$ (bottom to top). 
In (b) thick lines represent $\omega_\parallel(q)/\omega_\parallel(\infty)$ 
and thin lines represent $S(k)$ at $\tau_p=2\times 10^{-4}$ (solid)
and $2\times 10^{-2}$ (dashed). 
The inset in (b) shows the persistence time
dependence of $\omega_\parallel(\infty)\tau_p$. 
In (c) we show $\tau_\alpha$ (circles, left axis)
and the peak value of $g(r)$ (triangles, right axis), 
obtained from simulations.
In (d) we show $\tau_\alpha$ (circles) and the 
inverse self-diffusion coefficient, $0.1/D$ (squares) 
predicted by the theory.}
\end{figure}

In Fig.~\ref{fig2}(a-c) we show the dependence of $g(r)$, $S(q)$ and 
$\omega_\parallel(q)/\omega_\parallel(\infty)$ of the
self-propelled LJ system on  the persistence time $\tau_p$ 
at constant $T_{\text{eff}}=0.9$ (this is the lowest $T_{\text{eff}}$ 
at which we were able to simulate our one-component system
without observing spontaneous ordering). 
Again, the structure becomes progressively 
more pronounced as $\tau_p$ increases. Concurrently, correlations of 
particle velocities also develop, as revealed by $\omega_\parallel(q)/
\omega_\parallel(\infty)$. Finally, the quantity 
$\omega_\parallel(\infty) \tau_p$, which is a measure
of local mobility in the interacting self-propelled system, decreases. 

Because it incorporates these different trends, our theory
is able to account for the non-monotonic dependence of the 
dynamics on the persistence time. In particular, with increasing $\tau_p$ 
both $S(q)$ and $\omega_{\parallel}(q)$ grow for $q$ around $2\pi/\sigma$ and 
as a result $\mathcal{C}(q)$ gets smaller than $c(q)$, 
which is the likely source
of the speed-up in the dynamics. At larger $\tau_p$ decreasing 
$\omega_\parallel(\infty) \tau_p$
and increasingly more pronounced 
local structure seem to prevail upon the increase in velocity correlations 
resulting in the slowing down in the dynamics.   
In panels (c,d), we show the dynamics predicted by
the theory and obtained from simulations, respectively.
Clearly, our theory qualitatively predicts the 
non-monotonic dependence of the dynamics on the persistence time,
suggesting that including non-equilibrium velocity 
correlations $\omega_{\parallel}(q)$ in the theory is of major importance. 
Mode-coupling theories overestimate the slowing down of the 
dynamics~\cite{Goetzebook}, and this explains why our theory
predicts a more pronounced non-monotonic effect than in the simulations.
Describing quantitatively glassy dynamics at thermal equilibrium is a
notoriously difficult and open challenge \cite{BerthierBiroli}. 
Therefore, quantitative agreement should not be expected in the far 
from equilibrium context of self-propelled particles.

Various kinds of generalizations of mode-coupling theory for 
driven glassy fluids have been 
proposed~\cite{BerthierKurchan,Farage,Sperl,BBK,Reichman,Fuchs}.
In particular, Ref.~\cite{Farage} developed a theory for active
Brownian particles, where self-propulsion is added to a thermal
Brownian system, but this work differs from our approach on 
important aspects. First, correlations 
between positions and self-propulsions were neglected in \cite{Farage}.
Technically, this amounts to replacing the local steady-state average in 
Eq.~(\ref{Omega2}) by the average over the distribution of 
self-propulsions. As a result, 
$\left(\left<\mathbf{f}_i \mathbf{f}_j\right>_{\text{lss}} 
- \left<\mathbf{f}_i\right>_{\text{lss}} \left<\mathbf{f}_j
\right>_{\text{lss}}\right)$
gets replaced by $D_f \tau_p \delta_{ij}\boldsymbol{I}$, 
and the steady state correlation function $\omega_{\parallel}(q)$, 
which we have shown 
to play a central role, does not appear. 
Second, in our derivation 
we use projection operators defined in terms
of the steady state distribution, whereas Ref.~\cite{Farage}
uses the equilibrium distribution. As a result, 
the memory function derived in Ref.~\cite{Farage}
is the same as in the equilibrium mode-coupling theory while
ours is different. Because we consider an intrinsically nonequilibrium system,
there is no equilibrium distribution that we could use. Physically, 
the steady state distribution seems more natural since we
are describing fluctuations in the steady state. An obvious 
disadvantage of our 
choice is that we have to obtain the steady state correlation functions,
$S(q)$ and $\omega_{\parallel}(q)$, from simulations.

\section{Summary}

Using computer simulations and tools from liquid state theory 
we developed and analyzed 
an athermal system of interacting self-propelled particles. 
We showed that the microscopic structure and long-time dynamics
evolve nontrivially with the degree of 
nonequilibrium, which challenges equilibrium theories for dense fluids. 
We presented a theory for the collective dynamics of an active 
many-body system that can qualitatively 
capture these phenomena. In particular, the speed up 
of the dynamics of the active fluid was linked to emerging 
steady state correlations
of the velocities, an object with no relevant equilibrium counterpart. 
In future work, we will numerically characterize the 
approach to the glass transition in more detail for 
such nonequilibrium conditions.  
On the theory side, we will analyze the nature of the `nonequilibrium
glass transition'~\cite{BerthierKurchan} predicted by 
our theory. We shall also compare the dynamics  
predicted by the theory with that obtained from simulations at larger 
degrees of supercooling, and the relation between the ergodicity breaking
temperature predicted by our theory with the 
transition temperature obtained from simulations. 
To perform detailed quantitative 
comparisons, we will need to generalize the present 
results to binary mixtures. More generally, our work paves the 
way for developing a general, microscopic understanding of the
glassy dynamics of active materials when different interparticle 
interactions, self-propulsion mechanisms and possibly thermal noise are at play. 

\acknowledgments

This work started when GS was visiting Laboratoire Charles Coulomb of 
Universit\'{e} de Montpellier. The research in Montpellier is supported 
by funding from the European Research Council under the European
Union's Seventh Framework Programme (FP7/2007-2013) / 
ERC Grant agreement No 306845. GS and EF gratefully 
acknowledge the support of NSF Grant No.~CHE 1213401. 

\appendix

\section{Derivation of effective evolution operator 
$\Omega^{\text{eff}}(z)$}

In Appendix A we present an outline of a derivation of the effective  
$N$-particle evolution operator $\Omega^{\text{eff}}(z)$, Eq. (\ref{Omega2}). 
There are three main assumptions used in this derivation. First,
we assume that systematic currents vanish in the steady state of our system.
Second, we assume a separation of timescales for the structural relaxation and the
relaxation of the self-propulsions. Third, we approximate the dynamics in 
the space orthogonal to the local equilibrium distribution of self-propulsions 
by the free relaxation of the self-propulsions.

Equations of motion (\ref{eompos}-\ref{eomsp}) are equivalent to the 
following evolution equation for the $N$-particle joint distribution of 
positions and self-propulsions,
\begin{eqnarray}\label{Npeveq}
\partial_t P_N(\mathbf{r}_1,\mathbf{f}_1,...,\mathbf{r}_N,\mathbf{f}_N;t) =
\Omega P_N(\mathbf{r}_1,\mathbf{f}_1,...,\mathbf{r}_N,\mathbf{f}_N;t)
\nonumber \\
\end{eqnarray}
where $\Omega$ is the $N$-particle evolution operator given by Eq. (\ref{Omega1}).

We assume that the evolution equation (\ref{Npeveq}) has a steady state solution
$P_N^{\text{ss}}(\mathbf{r}_1,\mathbf{f}_1,...,\mathbf{r}_N,\mathbf{f}_N)$, and thus
\begin{equation}\label{ss}
\Omega P_N^{\text{ss}}(\mathbf{r}_1,\mathbf{f}_1,...,\mathbf{r}_N,\mathbf{f}_N) = 0.
\end{equation}
In the main text and in the following we use brackets 
$\left<\dots\right>$ to denote averaging over the joint steady state distribution 
of positions and self-propulsions, 
$P_N^{\text{ss}}(\mathbf{r}_1,\mathbf{f}_1,...,\mathbf{r}_N,\mathbf{f}_N)$.

From the joint steady state distribution we can obtain a steady state distribution
of positions, $P_N^{\text{ss}}(\mathbf{r}_1,...,\mathbf{r}_N)$,
\begin{equation}
P_N^{\text{ss}}(\mathbf{r}_1,...,\mathbf{r}_N) = 
\int  d\mathbf{f}_1 ... d\mathbf{f}_N 
P_N^{\text{ss}}(\mathbf{r}_1,\mathbf{f}_1,...,\mathbf{r}_N,\mathbf{f}_N).
\end{equation}
In the main text and in the following we use brackets 
$\left<\dots\right>_{\mathbf{r}}$ to denote averaging over a steady state distribution 
of positions, $P_N^{\text{ss}}(\mathbf{r}_1,...,\mathbf{r}_N)$.

We assume that in the steady state there are no systematic currents. To make
this statement more precise we first define the
current density by integrating Eq. (\ref{Npeveq}) over
self-propulsions to get the following continuity equation,
\begin{equation}
\partial_t P_N(\mathbf{r}_1,...,\mathbf{r}_N;t) =
- \sum_{i} \boldsymbol{\nabla}_i\cdot \mathbf{j}_i(\mathbf{r}_1,...,\mathbf{r}_N;t),
\end{equation}
where $P_N(\mathbf{r}_1,...,\mathbf{r}_N;t)$ is the N-particle distribution
of positions, 
\begin{equation}
P_N(\mathbf{r}_1,...,\mathbf{r}_N;t) = 
\int  d\mathbf{f}_1 ... d\mathbf{f}_N 
P_N(\mathbf{r}_1,\mathbf{f}_1,...,\mathbf{r}_N,\mathbf{f}_N;t),
\end{equation}
and $\mathbf{j}_i(\mathbf{r}_1,...,\mathbf{r}_N;t)$ is the current density 
of particle $i$,
\begin{eqnarray}
\lefteqn{\mathbf{j}_i(\mathbf{r}_1,...,\mathbf{r}_N;t) =}
\\ \nonumber && 
\xi_0^{-1} \int  d\mathbf{f}_1 ... d\mathbf{f}_N 
\left(\mathbf{F}_i + \mathbf{f}_i \right)
P_N(\mathbf{r}_1,\mathbf{f}_1,...,\mathbf{r}_N,\mathbf{f}_N;t).
\end{eqnarray}
Our assumption of the absence of systematic currents in the steady state is implemented
as follows,
\begin{equation}\label{curvan}
\xi_0^{-1}\int  d\mathbf{f}_1 ... d\mathbf{f}_N \left[\mathbf{F}_i + \mathbf{f}_i \right]
P_N^{\text{ss}}(\mathbf{r}_1,\mathbf{f}_1,...,\mathbf{r}_N,\mathbf{f}_N) = 0.
\end{equation}
The above equality implies that the local steady state average of the self-propulsion
is equal to the negative of the force,
$\left<\mathbf{f}_i\right>_{\text{lss}} = - \mathbf{F}_i$,
where
the local steady state average is defined as
\begin{eqnarray}
\lefteqn{\left< \dots \right>_{\text{lss}} = }
\\ \nonumber && 
\frac{1}{P_N^{\text{ss}}(\mathbf{r}_1,...,\mathbf{r}_N)}
\int  d\mathbf{f}_1 ... d\mathbf{f}_N
\dots
P_N^{\text{ss}}(\mathbf{r}_1,\mathbf{f}_1,...,\mathbf{r}_N,\mathbf{f}_N).
\end{eqnarray}

We assume that the self-propulsions relax faster than the positions of the particles. 
This assumption is applicable for strongly interacting 
systems where structural relaxation is slowing down, 
whereas the evolution of the self-propulsions stays,
by definition, independent of intermolecular interactions. 
The separation of the timescales for the structural
and self-propulsions relaxations allows us to derive an approximate equation
of motion for the distribution of particle positions. 

We define the projection operator on a local equilibrium-like
distribution (\textit{i.e.} on a distribution in which self-propulsions have 
a steady-state distribution for a given sample of positions),
\begin{eqnarray}
\lefteqn{\mathcal{P}_{\text{lss}} 
P_N(\mathbf{r}_1,\mathbf{f}_1,...,\mathbf{r}_N,\mathbf{f}_N;t)
= }
\nonumber \\ && =
\frac{P_N^{\text{ss}}(\mathbf{r}_1,\mathbf{f}_1,...,\mathbf{r}_N,\mathbf{f}_N)}
{P_N^{\text{ss}}(\mathbf{r}_1,...,\mathbf{r}_N)}
\int  d\mathbf{f}_1 ... d\mathbf{f}_N
P_N(\mathbf{r}_1,\mathbf{f}_1,...,\mathbf{r}_N,\mathbf{f}_N;t)
\nonumber \\ 
&& = \frac{P_N^{\text{ss}}(\mathbf{r}_1,\mathbf{f}_1,...,\mathbf{r}_N,\mathbf{f}_N)}
{P_N^{\text{ss}}(\mathbf{r}_1,...,\mathbf{r}_N)} 
P_N(\mathbf{r}_1,...,\mathbf{r}_N;t).
\end{eqnarray}
Next, we define the orthogonal projection,
$\mathcal{Q}_{\text{lss}} = \mathcal{I} - \mathcal{P}_{\text{lss}}$,
and write down equations of motion for 
$\mathcal{P}_{\text{lss}} 
P_N(\mathbf{r}_1,\mathbf{f}_1,...,\mathbf{r}_N,\mathbf{f}_N;t)$ and
$\mathcal{Q}_{\text{lss}} 
P_N(\mathbf{r}_1,\mathbf{f}_1,...,\mathbf{r}_N,\mathbf{f}_N;t)$,
\begin{eqnarray}\label{PPeom}
\lefteqn{
\partial_t \mathcal{P}_{\text{lss}} 
P_N(\mathbf{r}_1,\mathbf{f}_1,...,\mathbf{r}_N,\mathbf{f}_N;t)=}
\nonumber \\ && 
\mathcal{P}_{\text{lss}} \Omega \mathcal{P}_{\text{lss}} 
P_N(\mathbf{r}_1,\mathbf{f}_1,...,\mathbf{r}_N,\mathbf{f}_N;t) 
\nonumber \\ && + 
\mathcal{P}_{\text{lss}} \Omega \mathcal{Q}_{\text{lss}} 
P_N(\mathbf{r}_1,\mathbf{f}_1,...,\mathbf{r}_N,\mathbf{f}_N;t),
\end{eqnarray}
\begin{eqnarray}\label{QPeom}
\lefteqn{
\partial_t \mathcal{Q}_{\text{lss}} 
P_N(\mathbf{r}_1,\mathbf{f}_1,...,\mathbf{r}_N,\mathbf{f}_N;t) = }
\nonumber \\ && 
\mathcal{Q}_{\text{lss}} \Omega \mathcal{P}_{\text{lss}} 
P_N(\mathbf{r}_1,\mathbf{f}_1,...,\mathbf{r}_N,\mathbf{f}_N;t) 
\nonumber \\ && 
+ \mathcal{Q}_{\text{lss}} \Omega \mathcal{Q}_{\text{lss}} 
P_N(\mathbf{r}_1,\mathbf{f}_1,...,\mathbf{r}_N,\mathbf{f}_N;t).
\end{eqnarray}

Since our final goal is to calculate the intermediate scattering function given
by Eq. (\ref{Fkt}), we can assume that 
$\mathcal{Q}_{\text{lss}}
P_N(\mathbf{r}_1,\mathbf{f}_1,...,\mathbf{r}_N,\mathbf{f}_N;t=0)=0$. 
Then we  can solve Eqs. (\ref{PPeom}-\ref{QPeom}) for 
the Laplace transform, $\mathcal{LT}$, of 
$\partial_t \mathcal{P}_{\text{lss}} 
P_N(\mathbf{r}_1,\mathbf{f}_1,...,\mathbf{r}_N,\mathbf{f}_N;t)$,
which is given by
\begin{widetext}
\begin{eqnarray}\label{proj1}
\mathcal{LT}\left[\partial_t \mathcal{P}_{\text{lss}} 
P_N(\mathbf{r}_1,\mathbf{f}_1,...,\mathbf{r}_N,\mathbf{f}_N;t)\right](z) = 
\left[  \mathcal{P}_{\text{lss}} \Omega \mathcal{P}_{\text{lss}} 
+ \mathcal{P}_{\text{lss}} \Omega \mathcal{Q}_{\text{lss}}
\frac{1}{z-\mathcal{Q}_{\text{lss}} \Omega \mathcal{Q}_{\text{lss}}} 
\mathcal{Q}_{\text{lss}}\Omega\mathcal{P}_{\text{lss}} \right]
\mathcal{P}_{\text{lss}} P_N(\mathbf{r}_1,\mathbf{f}_1,...,\mathbf{r}_N,\mathbf{f}_N;z).
\nonumber \\ 
\end{eqnarray}
Using the assumption that systematic currents vanish in the steady state, 
Eq. (\ref{curvan}), one can show that the first term
inside the brackets on right-hand-side of Eq. (\ref{proj1}) vanishes. 
Furthermore, one can show that
\begin{eqnarray}\label{right}
\mathcal{Q}_{\text{lss}}\Omega \mathcal{P}_{\text{lss}} P_N(z) =
- \xi_0^{-1} \sum_i 
\left(\mathbf{f}_i - \left<\mathbf{f}_i\right>_{\text{lss}}\right)
P_N^{\text{ss}}(\mathbf{r}_1,\mathbf{f}_1,...,\mathbf{r}_N,\mathbf{f}_N)
\cdot
\left[\boldsymbol{\nabla}_i \frac{P_N(\mathbf{r}_1,...,\mathbf{r}_N;z)}
{P_N^{\text{ss}}(\mathbf{r}_1,...,\mathbf{r}_N)}\right]
\end{eqnarray}
and
\begin{eqnarray}\label{left}
\mathcal{P}_{\text{lss}} \Omega \mathcal{Q}_{\text{lss}} ...  =  
- \frac{P_N^{\text{ss}}(\mathbf{r}_1,\mathbf{f}_1,...,\mathbf{r}_N,\mathbf{f}_N)}
{P_N^{\text{ss}}(\mathbf{r}_1,...,\mathbf{r}_N)} 
\xi_0^{-1} \sum_i \boldsymbol{\nabla}_i \cdot
\int  d\mathbf{f}_1 ... d\mathbf{f}_N
\left(\mathbf{f}_i - \left<\mathbf{f}_i\right>_{\text{lss}}\right) ...
\end{eqnarray}
\end{widetext}

We note that $\mathcal{Q}_{\text{lss}}\Omega\mathcal{Q}_{\text{lss}}$
describes evolution in the space orthogonal to the local steady state space.
We assume that this evolution is entirely due to the free relaxation of
the self-propulsions. Specifically, we assume that 
$\mathcal{Q}_{\text{lss}}\Omega\mathcal{Q}_{\text{lss}}$ can be approximated by 
$\sum_{i=1}^{N} \frac{\partial}{\partial \mathbf{f}_i}
\left( \tau_p^{-1} \mathbf{f}_i  + D_f \frac{\partial}{\partial \mathbf{f}_i}\right)$. 
We note that this approximation physically means that the relaxation rate 
of the self-propulsions is not renormalized by the interparticle interactions.
Combining the last approximation with Eqs. (\ref{right}-\ref{left}) we get the following
approximate equality
\begin{widetext}
\begin{eqnarray} 
\lefteqn{
\mathcal{P}_{\text{lss}} \Omega \mathcal{Q}_{\text{lss}} 
\left(z-\mathcal{Q}_{\text{lss}}\Omega\mathcal{Q}_{\text{lss}}\right)^{-1} 
\mathcal{Q}_{\text{lss}}\Omega \mathcal{P}_{\text{lss}} 
P_N(\mathbf{r}_1,\mathbf{f}_1,...,\mathbf{r}_N,\mathbf{f}_N;z)
\approx }
\nonumber \\ && 
\frac{P_N^{\text{ss}}(\mathbf{r}_1,\mathbf{f}_1,...,\mathbf{r}_N,\mathbf{f}_N)}
{P_N^{\text{ss}}(\mathbf{r}_1,...,\mathbf{r}_N)} 
\xi_0^{-2} \sum_i \boldsymbol{\nabla}_i \cdot
\int  d\mathbf{f}_1 ... d\mathbf{f}_N
\left(\mathbf{f}_i - \left<\mathbf{f}_i\right>_{\text{lss}}\right)
\nonumber \\ && \times
\left[z - \sum_{j=1}^{N} \frac{\partial}{\partial \mathbf{f}_j}
\left( \tau_p^{-1} \mathbf{f}_j  + D_f \frac{\partial}{\partial \mathbf{f}_j}\right)
\right]^{-1}
\sum_l
\left(\mathbf{f}_l - \left<\mathbf{f}_l\right>_{\text{lss}}\right)
P_N^{\text{ss}}(\mathbf{r}_1,\mathbf{f}_1,...,\mathbf{r}_N,\mathbf{f}_N)
\cdot
\left[\boldsymbol{\nabla}_l \frac{P_N(\mathbf{r}_1,...,\mathbf{r}_N;z)}
{P_N^{\text{ss}}(\mathbf{r}_1,...,\mathbf{r}_N)}\right].
\end{eqnarray}
Now, we expand $\left[z - \sum_{i=1}^{N} \frac{\partial}{\partial \mathbf{f}_i}
\left( \tau_p^{-1} \mathbf{f}_i  + D_f \frac{\partial}{\partial \mathbf{f}_i}\right)
\right]^{-1}$ and integrate by parts.  Finally, we integrate both sides of
the resulting equation over self-propulsions and get the following expression
for the Laplace transform of $\partial_t P_N(\mathbf{r}_1,...,\mathbf{r}_N;t)$
\begin{eqnarray}\label{proj2}
\mathcal{LT}
\left[\partial_t P_N(\mathbf{r}_1,...,\mathbf{r}_N;t)\right](z) =
\xi_0^{-2} \sum_{i,j} 
\boldsymbol{\nabla}_i \cdot \left(z+\tau_p^{-1}\right)^{-1}
\left(\left<\mathbf{f}_i \mathbf{f}_j\right>_{\text{lss}} 
- \left<\mathbf{f}_i\right>_{\text{lss}} \left<\mathbf{f}_j\right>_{\text{lss}}\right)
\cdot
\left[-\mathbf{F}_j^{\text{\text{eff}}}+\boldsymbol{\nabla}_j\right]
P_N(\mathbf{r}_1,...,\mathbf{r}_N;z).
\end{eqnarray}
\end{widetext}
The right-hand-side of Eq. (\ref{proj2}) defines the effective 
evolution operator $\Omega^{\text{eff}}(z)$ given by Eq. (\ref{Omega2}).

\section{Derivation of an approximate evolution equation for $F(q;t)$}

In Appendix B we present an outline of a derivation of an approximate evolution 
equation for the intermediate scattering function $F(q;t)$, 
Eqs. (\ref{mfe}-\ref{vertex}). The framework of the derivation is based
on that of the derivation of the mode-coupling theory: first, the Laplace 
transform of the time derivative of $F(q;t)$ is expressed in terms of the
frequency matrix and the reducible \cite{CHess,SL} memory matrix. Next,
the reducible memory matrix is expressed in terms of the irreducible one. 
Finally, an approximate expression for the irreducible memory matrix 
in terms of the intermediate scattering functions is derived. 

The Laplace transform of the intermediate scattering function 
defined in Eq. (\ref{Fkt}) reads
\begin{eqnarray}\label{Fktz}
F(q;z)&=&N^{-1}\left<n(\mathbf{q})\left(z-\Omega\right)^{-1} n(-\mathbf{q})
\right> 
\nonumber \\ &=& N^{-1}\left<n(\mathbf{q}) 
\left(z-\Omega^{\text{eff}}(z)\right)^{-1} 
n(-\mathbf{q})\right>_{\bf r}
\end{eqnarray}
where $n(\mathbf{q})  = \sum_j e^{-i\mathbf{q}\cdot\mathbf{r}_j}$ 
is the Fourier transform of the microscopic density and 
in the second equality we used the effective evolution operator derived
in the previous section of this supplementary material.

To derive an approximate evolution equation for intermediate scattering function 
$F(q;t)$ we first define a projection operator on the microscopic density
\begin{equation}\label{Pn}
\mathcal{P}_n = ... \left. n(-\mathbf{q})\right>_{\bf r}
\left<n(\mathbf{q})n(-\mathbf{q})\right>_{\bf r}^{-1}
\left< n(\mathbf{q}) ... \right. .
\end{equation}
We should emphasize that projection operator $\mathcal{P}_n$ is defined in terms
of the steady state distribution, unlike in the approach of Farage and Brader \cite{Farage}. 
Next, we use the identity
\begin{eqnarray}\label{identity}
\frac{1}{z-\Omega^{\text{eff}}(z)} &=& 
\frac{1}{z - \Omega^{\text{eff}}(z)\mathcal{Q}_n} 
\\ \nonumber && + 
\frac{1}{z - \Omega^{\text{eff}}(z)\mathcal{Q}_n}\Omega^{\text{eff}}(z)\mathcal{P} 
\frac{1}{z-\Omega^{\text{eff}}(z)},
\end{eqnarray} 
where $\mathcal{Q}_n$ is the projection on the space orthogonal to that 
spanned by the microscopic density, 
$\mathcal{Q}_n = \mathcal{I} - \mathcal{P}_n$,
to rewrite the Laplace transform of the time derivative of $N F(q;t)$ as follows
\begin{widetext}
\begin{eqnarray}\label{timeder}
&& \mathcal{LT}[\partial_t N F(q;t)](z)= 
\left<n(\mathbf{q}) \Omega^{\text{eff}}(z) 
\frac{1}{z-\Omega^{\text{eff}}(z)} n(-\mathbf{q}) \right>_{\bf r} 
= \left<n(\mathbf{q}) \Omega^{\text{eff}}(z) \mathcal{P}_n 
\frac{1}{z-\Omega^{\text{eff}}(z)} n(-\mathbf{q}) \right>_{\bf r}
\nonumber \\ && + \left<n(\mathbf{q}) \Omega^{\text{eff}}(z) \mathcal{Q}_n 
\frac{1}{z-\Omega^{\text{eff}}(z)} n(-\mathbf{q}) \right>_{\bf r} 
=
\left<n(\mathbf{q}) \Omega^{\text{eff}}(z) n(-\mathbf{q})\right>_{\bf r}
\left<n(\mathbf{q})n(-\mathbf{q})\right>_{\bf r}^{-1}
\left< n(\mathbf{q}) \frac{1}{z-\Omega^{\text{eff}}(z)} n(-\mathbf{q}) \right>_{\bf r} 
\nonumber \\ && + 
\left<n(\mathbf{q}) \Omega^{\text{eff}}(z) \mathcal{Q}_n 
\frac{1}{z - \mathcal{Q}_n\Omega^{\text{eff}}(z)\mathcal{Q}_n} 
\mathcal{Q}_n \Omega^{\text{eff}}(z) n(-\mathbf{q})\right>_{\bf r}
\left<n(\mathbf{q})n(-\mathbf{q})\right>_{\bf r}^{-1}
\left< n(\mathbf{q})\frac{1}{z-\Omega^{\text{eff}}(z)}
n(-\mathbf{q}) \right>_{\bf r}.
\end{eqnarray}
We identify the analogue of the frequency matrix, $\mathcal{H}(q;z)$,
\begin{eqnarray}\label{freqmat}
\left<n(\mathbf{q}) \Omega^{\text{eff}}(z) n(-\mathbf{q})\right>_{\bf r}
= - \frac{ \mathbf{q}\cdot
\left< \sum_{i,j}\left(\left<\mathbf{f}_i \mathbf{f}_j\right>_{\text{lss}} 
- \left<\mathbf{f}_i\right>_{\text{lss}} \left<\mathbf{f}_j\right>_{\text{lss}}\right) 
e^{-i\mathbf{q}\cdot\left(\mathbf{r}_i-\mathbf{r}_j\right)}\right> 
\cdot\mathbf{q}}
{\xi_0^2\left(z+\tau_p^{-1}\right)} =  
- q^2 N \frac{\omega_{\parallel}(q)}{\left(z+\tau_p^{-1}\right)}
= -q^2 N \mathcal{H}(q;z)
\end{eqnarray}
\end{widetext}
where $\omega_{\parallel}(q)$ is the function quantifying correlations of
the velocities of individual particles that was 
introduced below Eq. (\ref{mfe}),
\begin{eqnarray}
\lefteqn{ \omega_{\parallel}(q) = } 
\nonumber \\ && \frac{1}{N\xi_0^2}
\hat{\mathbf{q}}\cdot
\left< \sum_{i,j}\left(\left<\mathbf{f}_i \mathbf{f}_j\right>_{\text{lss}} 
- \left<\mathbf{f}_i\right>_{\text{lss}} \left<\mathbf{f}_j\right>_{\text{lss}}\right) 
e^{-i\mathbf{q}\cdot\left(\mathbf{r}_i-\mathbf{r}_j\right)}\right>_{\bf r}
\cdot\hat{\mathbf{q}} 
\nonumber \\ && \equiv
\hat{\mathbf{q}}\cdot
\left< \sum_{i,j}\left(\mathbf{f}_i + \mathbf{F}_i\right)
\left(\mathbf{f}_j + \mathbf{F}_j\right)
e^{-i\mathbf{q}\cdot\left(\mathbf{r}_i-\mathbf{r}_j\right)}\right>
\cdot\hat{\mathbf{q}}.
\end{eqnarray}
Note that 
$\left< \left< ... \right>_{\text{lss}}\right>_{\mathbf{r}}=\left< ... \right>$.
Furthermore, we identify the analogue of 
the reducible \cite{CHess,SL} memory matrix, $\mathcal{M}(q;z)$,
\begin{widetext}
\begin{eqnarray}\label{memfunred}
&&  
\left<n(\mathbf{q}) \Omega^{\text{eff}}(z) \mathcal{Q}_n 
\frac{1}{z - \mathcal{Q}_n\Omega^{\text{eff}}(z)\mathcal{Q}_n} 
\mathcal{Q}_n \Omega^{\text{eff}}(z) n(-\mathbf{q})\right>_{\mathbf{r}} =
\left(\xi_0^4\left(z+\tau_p^{-1}\right)^2\right)^{-1}
\mathbf{q}\cdot\left< \sum_{i,j} e^{-i\mathbf{q}\cdot\mathbf{r}_i}
\left(\left<\mathbf{f}_i \mathbf{f}_j\right>_{\text{lss}} 
- \left<\mathbf{f}_i\right>_{\text{lss}} 
\left<\mathbf{f}_j\right>_{\text{lss}}\right)\cdot
\right. \nonumber \\ && \times \left.
\left[-\boldsymbol{\nabla}_j + \mathbf{F}^{\text{ss}}_j\right]
\mathcal{Q}_n
\frac{1}{z - \mathcal{Q}_n\Omega^{\text{eff}}(z)\mathcal{Q}_n}
\mathcal{Q}_n \sum_{l,m} \boldsymbol{\nabla}_l \cdot
\left(\left<\mathbf{f}_l \mathbf{f}_m\right>_{\text{lss}} 
- \left<\mathbf{f}_l\right>_{\text{lss}} \left<\mathbf{f}_m\right>_{\text{lss}}\right)
e^{i\mathbf{q}\cdot\mathbf{r}_m} \right>_{\mathbf{r}}\cdot\mathbf{q}
= q^2 N \mathcal{M}(q;z)
\end{eqnarray}
\end{widetext}

Following the procedure used previously to derive the mode-coupling theory
for Brownian systems we introduce irreducible memory matrix 
$\mathcal{M}^{\text{irr}}(q;z)$, which is given by the expression analogous to 
Eq. (\ref{memfunred}) but with the projected evolution operator
$\mathcal{Q}_n\Omega^{\text{eff}}(z)\mathcal{Q}_n$ replaced by 
irreducible evolution  operator $\Omega^{\mathrm{irr}}(z)$ \cite{CHess,SL,Kawasaki}. 
The relation between
$\mathcal{M}(q;z)$ and $\mathcal{M}^{\text{irr}}(q;z)$ reads
\begin{equation}\label{redirr}
\mathcal{M}(q;z)=\mathcal{M}^{\text{irr}}(q;z)-\mathcal{M}^{\text{irr}}(q;z)
\mathcal{H}^{-1}(q;z)\mathcal{M}(q;z).
\end{equation}

Combining Eqs. (\ref{timeder}-\ref{freqmat}) and (\ref{memfunred}-\ref{redirr})
we can write the Laplace transform of the time derivative of the 
intermediate scattering function in the following way,
\begin{eqnarray}\label{timeder2}
\lefteqn{
\mathcal{LT}[\partial_t F(q;t)](z)=}
\\ \nonumber &&  -q^2 \mathcal{H}(q;z)
\left(1+\mathcal{M}^{\text{irr}}(q;z)/\mathcal{H}(q;z)\right)^{-1} F(q;z)/S(q)
\end{eqnarray}
where $S(q)$ is the steady state structure factor, 
$S(q) = \left<n(-\mathbf{q})n(\mathbf{q})\right>_{\mathbf{r}}$. 
In turn, Eq. (\ref{timeder2}) can be re-written as
\begin{eqnarray}\label{timeder3}
&& \left(z+\tau_p^{-1}
+\left(z+\tau_p^{-1}\right)^2\mathcal{M}^{\text{irr}}(q;z)/\omega_{\parallel}(q)\right)
\\ \nonumber && \times 
\left(z F(q;z) - F(q;t=0)\right) = -\left(\omega_{\parallel}(q) q^2/S(q)\right)F(q;z).
\end{eqnarray}
Eq. (\ref{timeder3}) transformed back to the time domain reproduces 
Eq. (\ref{mfe}). In particular, the inverse Laplace transform of
$\left(z+\tau_p^{-1}\right)^2\mathcal{M}^{\text{irr}}(q;z)/\omega_{\parallel}(q)$
is equal to the irreducible memory function $M^{\text{irr}}(q;t)$ that enters into
Eq. (\ref{mfe}). 

To proceed we derive an approximate expression for the irreducible memory function 
$M^{\text{irr}}(q;t)$ in terms of the intermediate scattering functions. To this
end we first follow the steps of the derivation of the mode-coupling theory \cite{SL}
and replace projection operators $\mathcal{Q}_n$ in $M^{\text{irr}}(q;t)$ by 
projections on density pairs, and next factorize \emph{both} steady state
and time-dependent four-point correlations. As a result we get the following
approximate expression for $M^{\text{irr}}(q;t)$,
\begin{widetext}
\begin{eqnarray}\label{fac1}
M^{\text{irr}}(q;t) &\approx &
\frac{1}{N\xi_0^4\omega_{\parallel}(q)} \sum_{\mathbf{q}_1, ..., \mathbf{q}_8}
\hat{\mathbf{q}}\cdot\left< \sum_{i,j} e^{-i\mathbf{q}\cdot\mathbf{r}_i}
\left(\left<\mathbf{f}_i \mathbf{f}_j\right>_{lss} 
- \left<\mathbf{f}_i\right>_{lss} \left<\mathbf{f}_j\right>_{lss}\right)\cdot
\left[-\nabla_j + \mathbf{F}^{ss}_j\right]
\mathcal{Q}_nn_2(-\mathbf{q}_1,-\mathbf{q}_2)\right>_{\mathbf{r}}
\nonumber \\ && \times
\frac{\delta_{\mathbf{q}_1\mathbf{q}_3}\delta_{\mathbf{q}_2\mathbf{q}_4}
+ \delta_{\mathbf{q}_1\mathbf{q}_4}\delta_{\mathbf{q}_2\mathbf{q}_3}}
{N^2 S(q_1) S(q_2)}
N^2 F(q_3;t)F(q_4;t)\left(
\delta_{\mathbf{q}_3\mathbf{q}_5}\delta_{\mathbf{q}_4\mathbf{q}_6}
+ \delta_{\mathbf{q}_3\mathbf{q}_6}\delta_{\mathbf{q}_4\mathbf{q}_5}\right)
\nonumber \\ && \times
\frac{\delta_{\mathbf{q}_5\mathbf{q}_7}\delta_{\mathbf{q}_6\mathbf{q}_8}
+ \delta_{\mathbf{q}_5\mathbf{q}_8}\delta_{\mathbf{q}_6\mathbf{q}_7}}
{N^2 S(q_5) S(q_6)}
\left< n_2(\mathbf{q}_7,\mathbf{q}_8)
\mathcal{Q}_n \sum_{k,l} \nabla_k \cdot
\left(\left<\mathbf{f}_k \mathbf{f}_l\right>_{lss} 
- \left<\mathbf{f}_k\right>_{lss} \left<\mathbf{f}_l\right>_{lss}\right)
e^{i\mathbf{q}\cdot\mathbf{r}_l} \right>_{\mathbf{r}}\cdot\hat{\mathbf{q}}.
\end{eqnarray}
\end{widetext}
Next, we approximate the vertex functions. 
The justification for the form of this last approximation
will be discussed elsewhere. Here we present the approximate expression 
for the left vertex,
\begin{widetext}
\begin{eqnarray}\label{vertexex}
&& \xi_0^{-2} \hat{\mathbf{q}}\cdot\left< \sum_{i,j} e^{-i\mathbf{q}\cdot\mathbf{r}_i}
\left(\left<\mathbf{f}_i \mathbf{f}_j\right>_{lss} 
- \left<\mathbf{f}_i\right>_{lss} \left<\mathbf{f}_j\right>_{lss}\right)\cdot
\left[-\nabla_j + \mathbf{F}^{ss}_j\right]
\mathcal{Q}_n n_2(-\mathbf{q}_1,-\mathbf{q}_2)\right>_{\mathbf{r}}
\nonumber \\ &\approx & 
- i\mathbf{q}\cdot\mathbf{q}_1 N \left(\omega_{\parallel}(q)
\frac{1}{\omega_{\parallel}(\infty)}
\omega_{\parallel}(q_1)S(q_2) - \omega_{\parallel}(q)S(q_1)S(q_2)\right)
\delta_{\mathbf{q},\mathbf{q}_1+\mathbf{q}_2}
\nonumber \\ && 
- i\mathbf{q}\cdot\mathbf{q}_2 N \left(\omega_{\parallel}(q)
\frac{1}{\omega_{\parallel}(\infty)}
\omega_{\parallel}(q_2)S(q_1) - \omega_{\parallel}(q)S(q_1)S(q_2)\right)
\delta_{\mathbf{q},\mathbf{q}_1+\mathbf{q}_2},
\end{eqnarray}
\end{widetext}
where $\omega_{\parallel}(\infty)=\lim_{q\to\infty}\omega_{\parallel}(q)
\equiv \left(3N\xi_0^2\right)^{-1} 
\left<\sum_i \left(\mathbf{f}_i+\mathbf{F}_i\right)^2 \right>$.

Finally, we use Eq. (\ref{vertex}) and an analogous approximation for the right vertex
in Eq. (\ref{fac1}) and, after taking the thermodynamic limit, we obtain
Eqs. (\ref{memfction}-\ref{vertex}).


\begin{thebibliography}{99}
\bibitem{commentT}
Comparing systems at the same $T_{\text{eff}}$ is meaningful 
as these systems all have 
the same long-time dynamics in the absence of interactions. Note,
however, that for a given $T_{\text{eff}}$ the mean-square self-propulsion
acting on a given particle decreases with increasing $\tau_p$, 
$\left<\mathbf{f}_i^2\right> = 3 T_{\text{eff}} \xi_0/\tau_p$.
\bibitem{commentFs} 
A theory for the self-intermediate scattering function $F_s(q;t)$ 
can be obtained by extending the theory for $F(k;t)$ in the same way as in 
the mode-coupling approach \cite{Goetzebook}.
\bibitem{commentv} 
For our overdamped system 
$\xi_0^{-1}\left(\mathbf{F}_i+\mathbf{f}_i\right)$ is the velocity of the $i$th
particle, Eq.~(\ref{eompos}).
\bibitem{commentLJ} 
We have chosen the LJ interaction in order 
to simulate a system as similar to the KA binary mixture as possible. 
All results are presented in the standard LJ units. We used  
number density $\rho=1.02$ because at this density the 
long-time self-diffusion 
coefficients for the KA binary mixture and the one-component LJ system
near the `optimal' $\tau_p$ are very close.
\end{thebibliography}
\end{document}